# Enhancing the Guidance of the Intentional Model "MAP": Graph Theory Application

Rébecca Deneckère, Elena Kornyshova, Colette Rolland
Centre de Recherche en Informatique
Université Paris 1 Panthéon Sorbonne
Paris, France
rebecca.deneckere@univ-paris.fr,elena.kornyshova@malix.univ-paris.fr,colette.rolland@univ-paris.fr

*Abstract*— **The MAP model was introduced in information system engineering in order to model processes on a flexible way. The intentional level of this model helps an engineer to execute a process with a strong relationship to the situation of the project at hand. In the literature, attempts for having a practical use of maps are not numerous. Our aim is to enhance the guidance mechanisms of the process execution by reusing graph algorithms. After clarifying the existing relationship between graphs and maps, we improve the MAP model by adding qualitative criteria. We then offer a way to express maps with graphs and propose to use Graph theory algorithms to offer an automatic guidance of the map. We illustrate our proposal by an example and discuss its limitations.**

*Index Terms*— **MAP, Intentional Modeling, Graph.**

## I. INTRODUCTION

Prescriptive process models have been developed to bring order and structure to the software development process. However, these models are often too rigid and it is necessary to bring a bit of flexibility into them. The flexibility in process model will allow the engineer to adapt the process following the project at hand.

One attempt to enhance the process model flexibility has been made with the MAP model. MAP has been introduced by Rolland in the nineties in the field of Information System (IS) Engineering [1, 2, 3] and validated in several fields, either requirement engineering [4], method engineering [5] or process modelling [1]. This model introduces an intentional level into process modeling. This level is used to guide the engineer through the processes by dynamic choices of the tasks sequences. Each time that an intention is reached (a task executed), the model suggests the tasks that can be executed on the next step. As a result, the concrete process is not rigid but constructed dynamically following the situation.

This intentional level let us classify the MAP model as an intention-oriented language. As process models, maps can be compared to the various types of process modelling languages and formalisms [6]. They can be roughly classified according to their orientation to activity-sequence oriented languages (e.g., UML Activity Diagram [7]), agent-oriented languages (e.g., Role-Activity Diagram [8]) or state-based languages (e.g. UML state charts [7]). Most of these process models do not employ a goal construct as an integral part of the model. They use an internal view of a process, focusing on *how* the process is performed and externalizing *what* the process is intended to accomplish in the goal [9].

On the contrary, intention-oriented process modeling focuses on *what* the process is intended to achieve, thus providing the process rationale, i.e. *why* the process is performed. As a consequence, intentions to be accomplished are explicitly represented in the process model together with the different alternatives ways for achieving them [10]. It offers a new vision of IS process modelling by adding an intentional level. This level helps the engineer execute a process with a strong relationship to the situation of the project at hand.

Some works has been done to combine the MAP model with another kind of modeling, in order to enhance the practical use of maps. For instance, in [4], the authors offered a way to transform requirements represented in a map into a Data Flow Diagram. Thereafter, the construction design capability of the DFD is available for system implementation. They conclude that, even if the notation is about the same, the two diagrams are sufficiently different from one another as they address different views of a system. In a similar work [10], the authors tried to combine the intention-oriented modeling of maps with the formal state-based modeling of Generic Process Models. This led them to provide a state-based formalization of a map which allows its analysis and verification.

Despite these attempts, the MAP model lacks works on the automatic guidance of processes. This is mostly due to the difficulties of map guidance on the operational level which must allow using maps in order to execute processes with a narrow link with the intentional level.

We foresee enhancing the maps guidance on the operational level by their expression in terms of graphs and by adding valuations in the MAP model. This will offer a process execution associated to the map with a control of the map navigation.

The Graph theory offers a lot of techniques to be used on graphs. For instance, the shortest path problem is the problem of finding a path between two vertices such that the sum of the weights of its constituent edges is minimized. On a valuated map, it may be useful to find a path through the map which will satisfy some requirements of the application engineer (on time, cost and so on). The execution of the map will then be more flexible as the engineer will have the possibility to change the weight values of the map sections following the project at hand.

In addition, there is confusion between maps and graphs. Of course, the maps are visually constructed as graphs but their use is completely different. People who never studied the MAP model don't understand these differences, which are semantically based on the two different levels used: the intentional level of the MAP model and the operational level of the graphs.

Consequently, our aim in this work is (i) to offer an improvement of the MAP model with the addition of qualitative criteria in order to enhance the guidance through the maps and (ii) to propose a possible mapping between maps and graphs in order to use all the graphs techniques already defined in the literature.

The paper is organized as follows. Section II offers some theoretical background with an explanation of the MAP and graph models. Section III proposes an architecture to transform a map into a graph, with an algorithm using the two identified levels: intentional and operational. An illustration is given in section IV with an example. We consider the limitations of our proposal in Section V. Finally, the last section provides some conclusions and outlines future works.

## II. THEORETICAL BACKGROUND

### A. Intentional Model MAP

In order to define the MAP model, we must introduce the three levels of process representation: intentional, operational and executable. The *intentional level* is the level where the goals are defined and allowing the choice of alternatives following the situation at hand. The *operational level* represents the techniques to combine the choices made at the higher level. The *executable level* is the part of the process which realizes the goals (by the execution of guidelines, workflows...).

The MAP model [1, 11] allows specifying process models in a flexible way by focusing on the process intentions, and on the various ways to achieve each of these intentions. A map is presented as a diagram where nodes are *intentions* and edges are *strategies*. The directed nature of this diagram shows which intentions should precede which ones. Therefore, it is not imposed that once an intention is achieved the intention that immediately follows is directly undertaken. An edge enters a node if its associated strategy can be used to achieve the target intention (the given node). Since there can be multiple edges entering a node; a map is able to represent the many ways for achieving an intention.

The following figure shows the structure of a map (with UML [12] formalism).

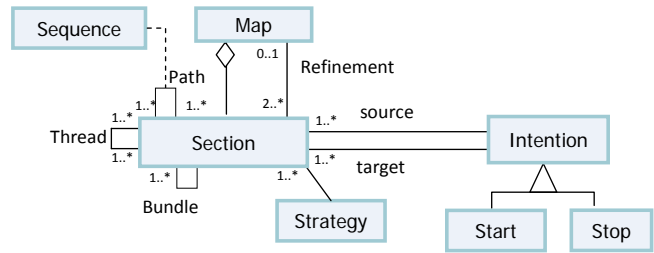

**Fig. 2. MAP model.**

A map (Figure 2) includes two predefined intentions: "Start" and "Stop", which mean accordingly the beginning and the end of the process. An important notion in process maps are the *sections* which represent the knowledge encapsulated in a triplet <source intention, strategy, target intention>, in other terms, the knowledge corresponding to a particular process step to achieve an intention (the target intention) from a specific situation (the source intention) following a particular technique (the strategy).

A specific manner to achieve an intention is captured in a map section whereas all sections having the same source and target intentions represent all the different strategies that may be used to achieve this target intention. In the same way, there may be several sections with the same source intention but different target ones. These ones show all the intentions that can be reached after the realisation of the source intention.

There are three possible relationships between sections namely the *thread*, *path* and *bundle* which generate *multi-thread* and *multi-path* topologies in a map [4].

A thread relationship shows the possibility for a target intention to be achieved from a source intention in many different ways. Each of these ways is expressed as a section in the map. Such a MAP topology is called a *multi-thread* and the sections participating in the multi-thread are said to be in a *thread relationship* with one another.

A path relationship establishes a precedence relationship between sections. For a section to succeed another, its source intention must be the target intention of the preceding one.

A bundle relationship shows the possibility for several sections having the same source and target intentions to be mutually exclusive.

A refinement relationship shows that a section of a map can be refined as another map through it. Refinement is an abstraction mechanism by which a complex assembly of sections at level *i+1* is viewed as a unique section at level *i*.

A map is a navigational structure as it allows the engineer to travel from Start to Stop. A map contains a finite number of *paths*, each of them prescribing a way to develop the product (each of them is a process model). No path is 'recommended' *a priori* as the engineer constructs his own path following the situation at hand. As a result, the MAP process model allows the development processes to be intention-oriented. At any moment, the application engineer has an intention, a goal in mind that he/she wants to fulfill.

Each section is then realized with the execution of a service. This service may be of different natures: a workflow, an algorithm, an intention achievement guideline (IAG) [1]… It allows guiding the application engineer in achieving an intention in a given situation in order to obtain the desired product. A section may also be refined in another map.

The following figure shows a map example.

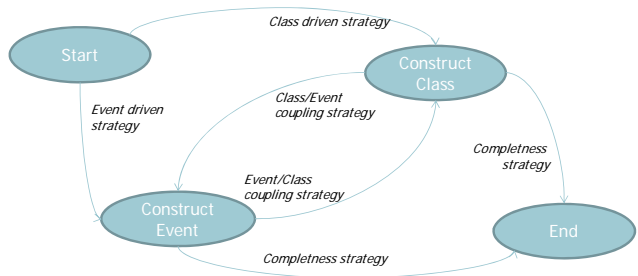

**Fig. 3. Map example.**

Note that the central part of the model, the section, is not represented on the maps with a unique symbol but with a set of concepts (source intention, target intention and strategy). As a matter of fact, the section concept doesn't exist as a symbol on a map. When reading a map, the executable service is then represented with two nodes and a vertex.

*B. Graphs*

Graph theory was born to study problems such as how to visit some places only once on a walk [13]. In mathematics and computer science, graph theory is the study of graphs: mathematical structures used to model pairwise relations between objects from a certain collection.
The following figure (Figure 2) shows the graph structure model used in this work.

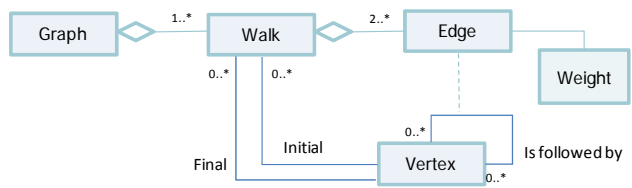

**Fig. 2. Meta model of Graph.**

Graphs are represented graphically by drawing a dot for every vertex, and drawing an arc between two vertices if they are connected by an edge. If the graph is directed, the direction is indicated by drawing an arrow, as shown in the figure 4.

A graph can be thought of as $G=(V,E)$, where $V$ and $E$ are disjoint finite sets. We call $V$ the *vertex set* and $E$ the *edge set* of $G$ [14, 15]. A *walk* is an alternating sequence of vertices and edges. An example of a walk is given on the preceding figure as this graph offers three different walks to go from the vertex 1 to vertex 4 (either directly or going through vertex 2 or through vertex 3). There is a distinction in graph theory between a *path* and a *walk* as a path is a walk with no repeated vertices. A walk begins with an initial vertex and ends with a final vertex.

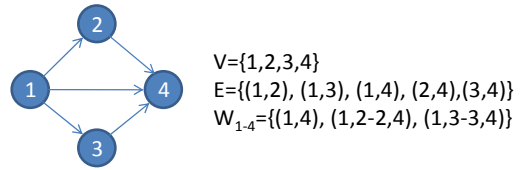

$V=\{1,2,3,4\}$
$E=\{(1,2), (1,3), (1,4), (2,4),(3,4)\}$
$W_{1-4}=\{(1,4), (1,2-2,4), (1,3-3,4)\}$

**Fig. 4. Graph example.**

A multigraph or pseudograph is a graph which is permitted to have multiple edges, (also called "parallel edges" [16]), that is, edges that have the same end nodes. Thus two vertices may be connected by more than one edge. Cycles are allowed in these graphs. A *cycle* is a path which ends at the vertex where it began.

A weighted graph associates a label (weight) with every edge in the graph.

### III. MAP EXPRESSION WITH GRAPHS

*A. Motivation*

MAP is a representation system that was originally developed to represent a process model expressed in intentional terms. However, there is no formal representation of this model allowing an easy way to automate its guidance. For its part, the graph theory has seen the development of algorithms to handle graphs which is of major interest to computer science. A 'concept-to-concept' comparison of these two models is presented in the appendix of this paper.

The use of these graph algorithms enhances the guidance of maps, especially as the MAP model is modified to manage specific weights. As a matter of fact, the navigation on the map is improved with the use of weights as the engineer makes decisions based on qualitative criteria. These criteria become a prerequisite to have a better guidance. Our first step is then to enhance the MAP model by integrating the weight concept in order to represent these criteria. Our second step is to use the graph algorithms within maps and we propose a correspondence between the two models.

*B. Enhanced MAP model with weight*

In order to improve the navigation on the maps and to offer a better guidance to the method engineer, a new concept has been added on the model which represents weight criteria affected to each section. The figure 5 shows the modified MAP model taking into account these concepts.

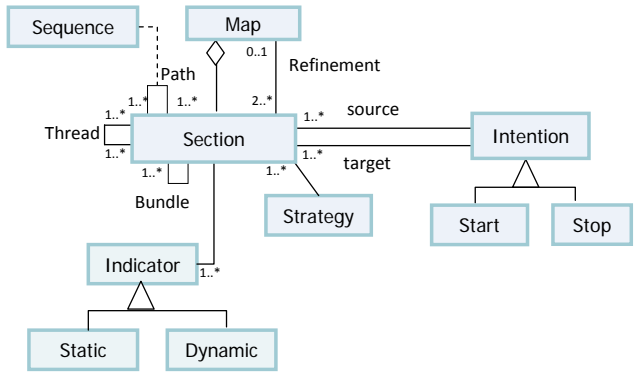

**Fig. 5. Enhanced MAP model.**

The weights are of two different types, either static or dynamic. The following table illustrates the indicators topology.

| Static indicators | | |
|---|---|---|
| Cost | Scale from 0 to 10 | Indicates the potential cost that the section realization will involve. |
| Time | Scale from 0 to 10 | Shows the time that the engineer will have to spend to realize the section. |
| ... | … | … |
| Dynamic indicators | | |
| Goal State | Scale from 0 to 10 | Gives an evaluation about the completeness of the intention realization. |
| Guideline Realization | Scale from 0 to 10 | Indicates the percentage of realization of the guideline corresponding to the section. |

Static indicators are weights that are evaluated in advance by the method engineer creating the map. They are usual criteria from project management, which contains evaluations of cost, time and so on [17]. For the sake of space, we restrain ourselves in this paper to the first two of them, as they are the most known in this field.

On the contrary, dynamic indicators are evaluated 'on the fly'. For instance, a section execution may completely realize an intention, which means that the Goal state indicator will have a value of 10. On the other hand, it may also incompletely realize it and a weight of only 5 over 10 may indicates that it is necessary to execute a cycle on the intention, in order to realize it more completely. [10] explores the use of combining the intention-oriented modeling of the map with a formal state-based modeling. The result shows a classification of the different cases of initial and final subsets of sections, taking into account the recursive cycles [10].

The Guideline Realization indicator gives an evaluation of the section completeness. Following the situation of the product in construction, the section may have to be executed several times in order to realize completely the guideline.

Let's take the following example. The map contains a section <Start, Initial identification strategy, Identify class> which will be realized by the following guideline <(Problem statement, Identify class by initial identification>. The problem statement is, for instance: 'The client may have several commands'. The realization of this guideline will first lead to the identification of the Client class. Even if the Goal state of the intention will be attained (as we have identified a class, which is our target intention), the guideline realization will not be complete as the problem statement contains another class to be identified: the Command class. As a result, the Goal State indicator will be equal to 10 but the Guideline realization indicator will be of 5 over 10. It will then be necessary to execute the guideline again in order to identify the second class.

### C. MAP and Graph correspondence

Transformation from map to graph is quite easy, as it is shown in the figure 6. To each map will correspond a graph (A). Each section of the map will then be transformed into graph vertices (B) and each identified section sequence will be shown as edges on the graph (C).

Finally, we may identify the correspondence between MAP indicator and graph weight (D). The section weights must then be evaluated in a single indicator that can be applied on the graph edges. The valuations of the graph edges are obtained based on aggregated values of sections: the section weight becomes the same valuation on all entering edges for a given section on the graph. We assume that each section has the same value independently of the previously realized section.

The MAP model allows refining a section with another map. This abstraction level is kept with the graphs as a node may also be refined as another graph.

These correspondences may be resumed in the Figure 6.

| Map concept | Graph concept | Correspondance |
|---|---|---|
| Map | Graph | Each map will be represented as a specific graph. |
| Section | Vertex | The executable service in a map is the section, which is represented as a vertex in a graph. |
| Sequence | Edge | The concept allowing the navigation in a map is the sequence of sections in the map, which may be represented as edges in a graph. |
| Indicator | Weight | The guidance parameters called indicators in the map are the weights in a graph. |
| Path | Walk | The set of section sequences (path) is clearly a graph walk. |
| Thread | Set of adjacent vertices | The thread is the possibility to attain a target intention from a source intention with several strategies. In a graph, the representation of this set of sections will be the set of vertices having two specific edges, one which shares the same start-vertex and the other the same end-vertex. |
| Bundle | Set of adjacent vertices | The thread is the possibility to attain a target intention from a source intention with several strategies but with an exclusive OR which means that only one of these sections may be used in the complete navigation. The representation on the graph is the same than |

|  |  | for the thread. |
|--|--|-----------------|

**Fig. 6. MAP and Graph Correspondences**

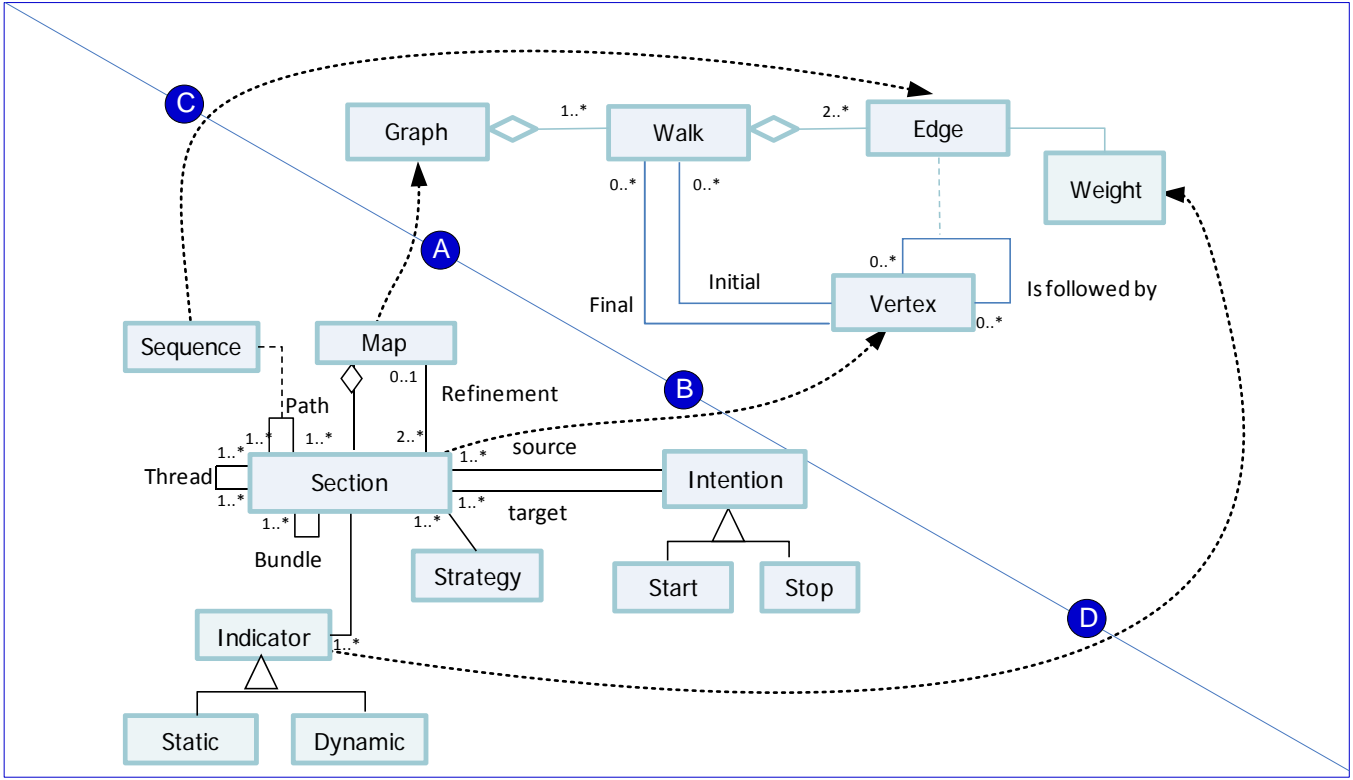

**Fig. 6. MAP and Graph Correspondence**

*D. Proposed "Architecture"*

The utilisation of a graph structure will allow using an operational level on our process model whereas the utilisation of a map adds the notion of an intentional model. These two levels may be used in a combined way to combine their own advantages. Firstly, the graph theory allows the application of a number of specific algorithms (complete path, shortest path…). Secondly, the MAP model allows an execution of a process on the fly, following the always permanently evolving situation of the product to help choosing the right objective to attain. Figure 7 illustrates the combined utilisation of the two levels.

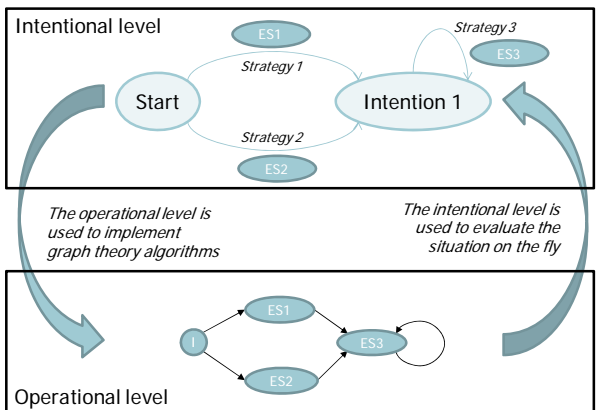

**Fig. 7. Illustration of the two levels combination**

The way of working is as follows. From an intention in the map, the engineer will use the graph operational level to apply a graph theory algorithm. This algorithm will propose a vertex. The engineer will then execute the corresponding section on the map intentional level (this section represents a specific executable service *ES*). This process is repeated until the engineer attains the end intention of the map. The proposed algorithm is detailed as follows.

Given a map M and a corresponding graph G (and all refined maps M' having a corresponding sub-graph G').

Given a graph theory algorithm the engineer wants to use to help the guidance of the MAP process.

Given I the initial node of the graph and F its final node.

```
Current := I
Num := 1
Path[1] := I
result:= Apply-algo (I, FV)
list:= (second-vertex(result))
next := list                                    ①

While next != F do                              ②

   Realize (next)                               ③

   If bundle(next)
    G=Delete-others-vertices-bundle(G, next)    ④
   end-if

   G=Evaluate-weight(G)                         ⑤

   Current := next
   Num := num+1
   Path[num] := current
   result:= Apply-algo (path[num], F)           ⑥
   list:= list + second-vertex(result)
   next := best-section (list)
End-while
```

Principal steps of this algorithm are explained as follows.
① After the initialization of several variables, we apply the chosen graph theory algorithm from the initial vertex to the final vertex in order to obtain a path. The second vertex of this path is then put in a list of the potential sections which can be attained at that point.
② The algorithm contains a loop (while) that will be performed until the engineer reaches the final intention of the map.
③ The engineer reaches the intentional level by realizing the vertex corresponding section. It is to note that the realization of this section may be reflexive, as it may be refined as another map, and we may have to restart this algorithm with the corresponding sub-graph G'.
④ If the realized section is a part of a bundle, we know that we will not be allowed to execute another section of this bundle. As a result, the corresponding vertices of the graph G are deleted to ensure this rule.
⑤ The execution of the section will modify the weights of the dynamic indicators (Goal state and Guideline realization). They are evaluated and integrated in each of the edges weight.
⑥ We can go back to the operational level and continue our path a little further, until we attain the final vertex. Note that the followed path is remembered in order to be able to go backward in the map. In order to know which sections are attainable from our intention, we apply the chosen graph theory algorithm from all realized sections (all vertices of the chosen path) to the final vertex in order to obtain the potential paths. The second vertex of each path is then put in a list of the potential sections which can be attained at that point. An evaluation is then made between all the elements of this list in order to choose between a backward or a forward path (this evaluation take into account the dynamic indicators in order to know if it is necessary to go backward in the map).

## IV. APPLICATION EXAMPLE

### A. Example Description

In order to illustrate our proposal, we have chosen a map describing the construction process of an O* model [18] (Figure 8).

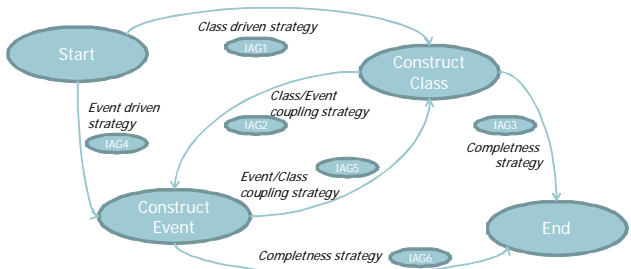

**Fig. 8. Map of the O* model construction process**

There are two different ways to initiate the construction of an O* model. On the one hand, the engineer may identify the classes of the model (by initial identification, by composition, by inheritance, by reference...) with the *Class driven strategy* (section *IAG1*). On the other hand, he/she may also choose to construct the model by the identification of events (by initial identification, by top-down or bottom up strategy...) with the *Event driven strategy* (section *IAG4*).

These two sections are refined by the execution of two strategic guidelines which are themselves represented as maps. For instance, the following Figure (Figure 9) represents the refined section <Start, Class Driven Strategy, Construct Class>.

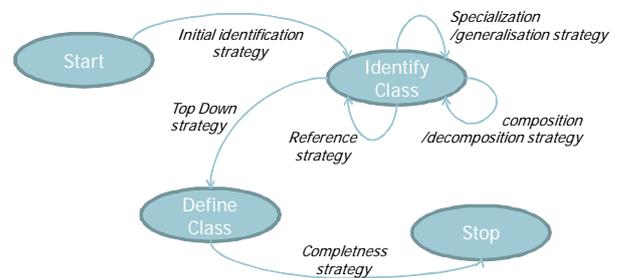

**Fig. 9. IAG1 section refined map**

When a class has been constructed, the engineer may then expand it, with the *Class/event coupling strategy* (section *IAG2*) or by the identification of related events - either the events that trigger operations of this class or a possible internal event (an event which is triggered by a specific modification of a class state). In the same way, the *Event/Class coupling strategy* (section *IAG5*) allows the engineer to identify related classes (either the class impacted by the operations triggered by the event or a class from which a state modification has triggered this particular event). These two sections also are refined by two other maps.

The *completeness strategy* allows the engineer to end the process with a verification of the obtained product model, either after having created a class (section *IAG3*) or an event (section *IAG6*). Note that the completeness will be obtained only if there have been a coupling of these concepts. The engineer will then not have the possibility to end the map without coupling the created events to the created classes (and vice versa).

The map creator has defined the following values for the weight indicators. The Goal state and Guideline Realization values are always equal to 0 before any execution of the process, as they are dynamic values which will be evaluated 'on the fly'.

| Section | Static Aggregated Value |
|---|---|
| Start-class driven strategy – construct class – IAG1 | 9 |
| Start – event driven strategy – construct event – IAG4 | 7 |
| Construct class – class/event coupling strategy – construct event – IAG2 | 6 |
| Construct event – event/class strategy – construct class – IAG 5 | 8 |
| Construct class – completeness strategy – end – IAG3 | 1 |
| Construct event – completeness strategy – end – IAG6 | 1 |

The static aggregated value of the section weights is calculated before any navigation on the map. It represents the complexity of the corresponding executable services. In this example, some sections may be refined by another maps which are themselves complex guidelines. The section weight represents the complexity of the map hierarchies.

*B. Map Transformation*

Based on rules presented in section III we describe the 0* map as a graph (Figure 10).

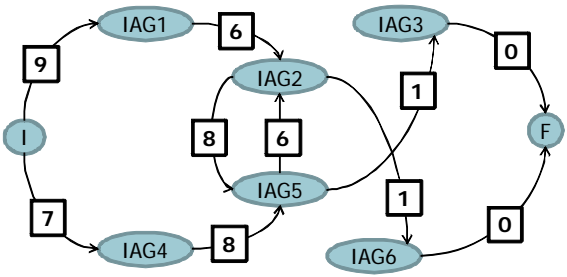

**Fig. 10. The weighted graph corresponding to the O* model construction process.**

Assume this graph is described in an interpretable language. There are different ways to store graphs in a computer system (either lists or matrices or both of them). For instance, the representation suggested by van Rossum, in which a hash table is used to associate each vertex with an array of adjacent vertices, can be seen as an instance of this type of representation [19]. Another possible way to represent this graph will be to use the XGMML (eXtensible Graph Markup and Modeling Language) structure [http://www.cs.rpi.edu/~puninj/XGMML]. We chose the matrix representation in this example.

Our example graph may be described with the following sets E and V.

```
V = {I,IAG1,IAG2,IAG3,IAG4,IAG5,IAG6,F}

E = {(I,IAG1,9), (I,IAG4,7), (IAG1, IAG2,6),
(IAG2,IAG5,8), (IAG2,IAG6,1), (IAG3,F,0),
(IAG4,IAG5,8), (IAG5,IAG2,6), (IAG5,IAG3,1),
(IAG6,F,0)}
```

This graph can be represented by the following Python data structure:

```
G = { 'I':    ['IAG1, 9', 'IAG4, 7'],
      'IAG1': ['IAG2, 6'],
      'IAG2': ['IAG5, 8', 'IAG6, 1'],
      'IAG3': ['F, 0'],
      'IAG4': ['IAG5, 8'],
      'IAG5': ['IAG2, 6', 'IAG3, 1'],
      'IAG6': ['F, 0'],
      'F':    []}
```

*C. Application of map enhanced by "Graph Layer"*

This section illustrates the use of this double view of a process with a simple example. We want to construct an O* model for the following description of the project: *"The client phones to the company to obtain a reservation. He gives information about himself (name, address, and phone) and about the reservation (beginning date, ending date). The client may also cancel an 'OK' reservation."*

The engineer chooses to apply systematically the algorithm of the *minimal weight path*. The procedure 'Apply-algo (A, B)' will then choose a path between the vertices A and B which will be the one with the minimal weight.

Initial affectations:
```
Current = I
Num = 1
Path[1]= I
result:= Apply-algo (I, F)
list:=(second-vertex(result))= IAG4
next := IAG4
```

First iteration:
```
Realize (IAG4)
G=Evaluate-weight(G)
Current := IAG4
Path[2] := IAG4
Result = apply-algo (IAG4, F)
second-vertex(result) = IAG5
list = (IAG4, IAG5)
Next = best-section (IAG4,IAG5) = IAG4
```

The first iteration of the algorithm will allow realizing the section IAG4 which will create the event 'Demand of reservation'.

After this iteration, the engineer analyzes the situation and establishes that the intention is realized at 50 %. The guideline of this section is not complete as there is still an event to identify.

Second iteration:
```
Realize (IAG4)
G=Evaluate-weight(G)
Current := IAG4
Path[3] := IAG4
Result = apply-algo (IAG4, F)
second-vertex(result) = IAG5
list = (IAG4, IAG5)
```

```
Next = best-section (IAG4, IAG5) = IAG5
```

The second iteration performs a second time the section IAG4, in order to create the event 'Cancel a reservation'. In this stage, the engineer establishes that the given intention is completed. Based on these findings, the engineer continues to apply the algorithm from IAG4 without return back to the initial vertex (I).

The algorithm application identifies that the next section to realize is IAG5, therefore it is necessary to couple found events to classes.

Third iteration:
```
Realize (IAG5)
G=Evaluate-weight(G)
Current := IAG5
Path[4] := IAG5
Result = apply-algo (IAG5, F)
second-vertex(result) = IAG3
list = (IAG4, IAG5, IAG3)
Next = best-section (IAG4,IAG5,IAG3) = IAG3
```

The third iteration will go further in the map in order to realize the section IAG5 to identify the classes coupled to the event 'Demand of reservation', which are firstly the reservation class, but also the client class (which we find by a study of the reference links of the reservation class).

The next section to realize is IAG3, to test the completeness of the product.

Fourth iteration:
```
Realize (IAG3)
G=Evaluate-weight(G)
Current := IAG3
Path[5] := IAG3
Result = apply-algo (IAG3, F)
second-vertex(result) = F
list = (IAG4, IAG5, IAG3, F)
Next = best-section (IAG4,IAG5,IAG3,F) = F
```

The fourth iteration performs the section IAG3 which tests the completeness of the desired product.

At the intentional level, the engineer reaches the *Stop* intention; therefore the map navigation is finished. At the operational one, the process gets into the final vertex F, which ends the navigation through the graph.

This example shows the use of the shortest past algorithm on a map expressed as a graph. The guidance has been improved to ease the engineer decisions as each section was proposed to him in the course of the process, by minimizing the process complexity.

## V. DISCUSSION

The main application that can be found with this work is an improvement of the guidance in the map, coupled with an automation of the navigation. The introduction of the dynamic criteria helps to create a better guidance.

The MAP model has already been used to show the variability of business process [20]. The use of algorithms from the graph theory field will help to identify, for instance, the number of possible paths which enter in the field of calculating process variability.

However, we cannot apply our approach for maps containing specific intentions maintaining a state that has already been reached [10]. In such isolated cases, recursive strategies are aimed at verifying that the desired state is not violated and the problem is that our algorithm will not be able to go out of this intention and will then repeat it indefinitely. As a consequence, the engineer has to decide to go further in the map without automatic guidance.

## VI. CONCLUSION

This paper offers a clarification of the two concepts of maps and graphs. We have highlighted the differences between the two models and offer a way to transform one into the other. This work proposes a possible operationalisation of the MAP model with a combination of an intentional level (map) and an operational level (graph). The use of the latter offers an opportunity to automate the guidance of the former with the help of an algorithm. This proposed algorithm goes from one level to the other after each section execution. These transformation and algorithm offer the possibility to use graph theory algorithms (dealing with directed valuated multigraphs) as required by the engineer on any valuated map.

Our future work is:

• to refine the indicator typology in order to express the criteria which will be important to engineers when guided through the map.

• to apply graph algorithms to maps in order to measure the variability of processes.

• to extend our proposal to other application domains, for instance, to dynamic workflows.

APPENDIX. COMPARISON BETWEEN GRAPHS AND MAPS

Despite similarities maps could not be considered as graphs. The following table illustrates the differences between the two types of graphics.

| Graph (simple graphs) | Graph (multigraph, pseudograph) | Map |
|---|---|---|
| **Structural** | | |
| No parallel edges | Parallel edges | Parallel strategies |
| No cycles | Cycles | Cycles |
| No loops | Loops | Loops |
| No backward | No backward without loop | Backward |
| Labeled vertex | Labeled vertex | Intention name |
| **Dynamical** | | |
| A vertex is used to show an incidence between two edges. | A vertex is used to show an incidence between two edges. | A strategy is labeled with an intentional objective. |
| Weight (labeled edges) | Weight (labeled edges) | No Weight, only the strategy name |

As described on the table above, a map cannot be used in the same way as a simple graph. First of all because simple graphs don't include parallel edges, cycles or loops, as multigraphs do. Moreover, even if we look more closely at the multigraph definition, the dynamic dimensions of the edges aren't the same. A graph edge is only used to link two vertices to allow the construction of paths (sequencing) whereas a map strategy adds a semantically richer dimension within an intentional level.

The backward issue is also an important one. No graphs allow to go back to a preceding vertex already visited (except if you have a loop which is the only case that will allow this possibility). On maps, the engineer has the possibility to go back to any intention already reached if he thinks it will obtain a better product that way.

Another difference is the weight value that may be added to the graph edges. The MAP model doesn't include values to differentiate between the different allowed strategies to realize an intention.

As a result, two edges linking the same vertices will not have the same signification on a map than on a graph.